\documentstyle[aps,prb,twocolumn,epsfig]{revtex}

\newcommand{\beq}{\begin{equation}}
\newcommand{\eeq}{\end{equation}}
\newcommand{\bea}{\begin{eqnarray}}
\newcommand{\eea}{\end{eqnarray}}

\draft
\begin{document}

\title{Symmetric Anderson impurity model with a narrow band}

\author{W. Hofstetter and S. Kehrein\cite{HU}}
\address{Theoretische Physik III, Elektronische Korrelationen und Magnetismus, 
Universit\"at Augsburg, D-86135 Augsburg, Germany}
\date{\today}

\maketitle

\begin{abstract}
The single channel Anderson impurity model is a standard model for the 
description of magnetic impurities in metallic systems. Usually, the bandwidth 
represents the largest energy scale of the problem. In this paper, we analyze 
the limit of a \emph{narrow} band, which is relevant for the 
Mott-Hubbard transition in infinite dimensions. For the symmetric model we 
discuss two different effects. (i) The impurity contribution to the density 
of states at the Fermi surface always turns out to be negative in such systems. 
This leads to a new crossover in the thermodynamic quantities that we investigate 
using the numerical renormalization group. (ii) Using the Lanczos method, we calculate 
the impurity spectral function and demonstrate the breakdown of the skeleton expansion 
on an intermediate energy scale. Luttinger's theorem, as an example of the local 
Fermi liquid property of the model, is shown to still be valid. 
\end{abstract}

\section{Introduction}
\noindent The single channel Anderson impurity model \cite{Anderson 61} is commonly used 
to describe the physics of a magnetic impurity in a conducting host and exhibits different 
types of behaviour such as mixed valency or the Kondo effect. 
Usually the case of a flat band is of interest, where the 
conduction electron bandwidth represents the largest energy scale in the problem.
For this particular case the model has been solved using the 
Bethe ansatz \cite{Wiegmann 81}, which enables one to calculate static properties, e.g. 
the impurity susceptibility or the specific heat. A numerical solution of the problem has been 
given by Krishna-Murthy et al. \cite{Krishnamurthy 80}.

In the context of the Hubbard model in high dimensions \cite{Metzner 89} 
(``dynamical mean field theory''), the physics of the correlated lattice system is 
described by a single Anderson impurity interacting with a bath whose properties are 
determined in a self-consistent way \cite{Georges 96}.   
The bath is characterized by a hybridization strength of the order of the bandwidth and  
a non-constant density of states. 
For this case no exact results exist and one either has to rely on analytical
approximations (which are, however, not available in the whole parameter space)
or numerical means.

In this paper we numerically solve the Anderson impurity model in the extreme limit where
the bandwidth is much smaller than the hybridization ({\em narrow band} systems).
In the context of the Mott-Hubbard transition in high
dimensions such models are self-consistently generated in one possible transition 
scenario \cite{Moeller 95}. By analyzing the narrow band limit (without the additional
complications caused by the $d\rightarrow\infty$ self-consistency condition) 
we will establish two characteristic new features of Anderson impurity models
that to some extent will also hold for intermediate situations: i) The impurity 
density of states $\rho_d(\omega)$ and the change in the density of states of 
the total system due to the impurity $\Delta\rho(\omega)$
show a very different behaviour since the conduction electrons react to the 
presence of the impurity. This is in contrast to the flat band case where always 
$\rho_d(\omega) = \Delta\rho(\omega)$. It leads to an interesting crossover in the 
thermodynamic impurity properties of the system when the interaction is turned on. 
ii) The skeleton expansion, which plays an important role in deriving properties
of the interacting system, breaks down on intermediate energy scales. This has
important implications for the Mott-Hubbard transition in $d=\infty$ as will be
explained below. 
 
\section{Density of states}
\noindent The Hamiltonian of the symmetric model is given by 
\bea \label{Hamiltonian}
H &=& \sum_{{\bf k},\mu} \epsilon_{\bf k} c^\dagger_{{\bf k}\mu} c^{\phantom{\dagger}}_{{\bf k}\mu} 
+ \sum_{{\bf k},\mu} (V^{\phantom{\dagger}} _{{\bf k}d} 
\,  c^\dagger_{{\bf k}\mu} d^{\phantom{\dagger}}_{\mu} + 
V^*_{{\bf k}d} \, d^\dagger_{\mu} c^{\phantom{\dagger}}_{{\bf k}\mu}) \nonumber \\
&&+U\left(d^\dagger_{\uparrow} d^{\phantom{\dagger}}_{\uparrow} - \frac{1}{2}\right)
\left(d^\dagger_{\downarrow} d^{\phantom{\dagger}}_{\downarrow} - \frac{1}{2}\right)
\eea
with the conduction electrons $c_{\bf k}$ and the impurity orbital $d$.
In the following, we will restrict ourselves to scalar $k's$, implying  - 
if necessary - a reduction to s-waves in the conduction band. 
The impurity density of states is defined by 
\beq
\rho_d(\omega) = -\frac{1}{\pi} Im\, G_{dd}(\omega^+) \ ,
\eeq
where $G_{dd}$ is the retarded zero temperature Green's function 
of the impurity orbital.
On the other hand, the total change of the density of states due to the  
introduction of the impurity into the conduction band is given by 
\bea \label{totaldos}
\Delta\rho(\omega) &=& -\frac{1}{\pi} Im\left\{\sum_k G_{kk}(\omega^+) 
+ G_{dd}(\omega^+)\right\} \nonumber \\
&&+\frac{1}{\pi} Im \sum_k G^{(0)}_{kk}(\omega^+) \ ,
\eea
where $G^{(0)}$ refers to the Green's function of the conduction electrons without the impurity.
Using the equations of motion for $G_{k k'}$ and $G_{dk}$ one finds 
\beq
\Delta\rho(\omega) = -\frac{1}{\pi} Im \left\{G_{dd}(\omega^+) 
\left(1-\frac{\partial}{\partial \omega} \sum_k \frac{V^2_k}{\omega^+ - \epsilon_k}\right)\right\} \ .
\eeq
Therefore the total change in the DOS can be expressed as a function of 
the impurity Green's function. 
We introduce the notation 
\beq
\sum_k\, \frac{V_k^2}{\omega^+ - \epsilon_k} = \Lambda(\omega) - i\Delta(\omega) \ ,
\eeq
where $\Delta(\omega)$ is the hybridization function and the real part is 
given by the principal value integral
$\Lambda(\omega) = {\mathcal{P}} \int \frac{d\epsilon}{\pi}
 \, \frac{\Delta(\epsilon)}{\omega - \epsilon} $.
Now we can express $\Delta\rho(\omega)$ in terms of the impurity density of states and write 
\beq \label{totaldosII}
\Delta\rho(\omega) = \rho_d(\omega)\,\left(1-\frac{\partial\Lambda}{\partial \omega}\right)
 - \frac{\partial \Delta}{\partial \omega} \, {\mathcal{P}} \int \frac{d\epsilon}{\pi} 
\frac{\rho_d(\epsilon)}{\omega - \epsilon} \ .
\eeq
We define the {\em narrow band} limit by the property
\beq
\left.\frac{\partial\Lambda(\omega)}{\partial\omega}\right|_{\omega=\epsilon_F}
\gg 1
\label{def_narrowband}
\eeq
leading to a negative coefficient multiplying the first term in
(\ref{totaldosII}) at the Fermi energy.

We will study one exemplary realization of a {\em narrow band} system
with a constant hybridization function $\Delta(\omega)=\Delta$ and
conduction band energies extending from $-D$ to $D$ (we set 
$\epsilon_F=0$ in the sequel). This leads to 
\beq
\left.\frac{\partial\Lambda(\omega)}{\partial\omega}\right|_{\omega=0}
=\frac{2\Delta}{\pi D}
\eeq
and hence the {\em narrow band} limit is defined by $\Delta\gg D$.
However, the main conclusions in the following analysis equally hold
for other realizations of narrow band systems (\ref{def_narrowband})
as well, in particular even for systems without band edges at all.

We notice that for our model the second term of (\ref{totaldosII}) vanishes 
inside the conduction band. 
Luttinger's theorem \cite{Luttinger 61,Langreth 66} for the symmetric Anderson model, 
which we will later verify numerically also for the narrow band case, 
ensures the ``pinning'' of the impurity spectral function 
at the Fermi energy at its noninteracting value
\beq \label{pinning}
\rho_d(0)=\frac{1}{\pi\Delta}\ .
\eeq 
Therefore the first term in (\ref{totaldosII}) gives a negative 
contribution
\beq \label{reduction}
\Delta\rho(0) = \frac{1}{\pi \Delta} \left(1- \frac{2 \Delta}{\pi D}\right)
\eeq
at the Fermi energy. In the following we discuss the relation of $\Delta\rho(\omega)$ to 
thermodynamic properties of the model.

\section{Calculation of the impurity susceptibility}
\begin{figure}[t]
\begin{center}
\epsfig{file=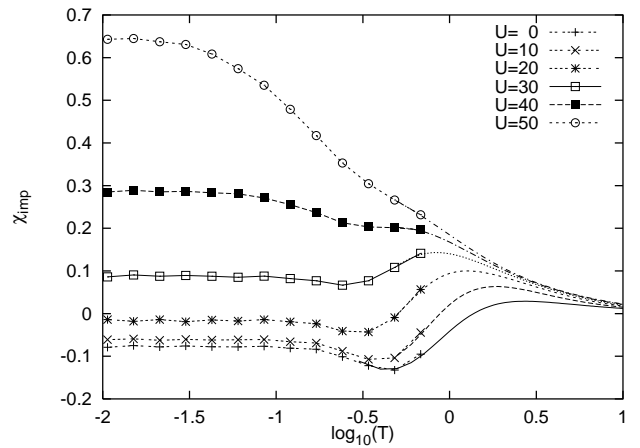,width=\linewidth}
\end{center}
\caption{\label{fig:chi_impurity}Impurity susceptibility vs.\ logarithmic temperature 
at hybridization $\Delta=10$ and half bandwidth $D=1$ for different interactions $U$. }
\end{figure}

In noninteracting fermionic models, the total density of states at the 
Fermi energy determines thermodynamic properties like the static susceptibility $\chi$.
Introducing an impurity into the system induces a change in $\chi$ proportional to $\Delta\rho(0)$.
For the narrow-band Anderson systems analyzed here one therefore expects   
a negative impurity contribution to the susceptibility. This is obviously true for
$U=0$ where one obtains the usual Pauli susceptibility in dimensionless units ($\mu_B = \hbar = g = 1$)
\beq   \label{Pauli}
\chi_{\mathrm{imp}} = \frac{\Delta\rho(0)}{2} = \frac{1}{2\pi \Delta}
\left(1- \frac{2 \Delta}{\pi D}\right)  \ .
\eeq
In order to determine whether this holds also for the interacting case, 
we have calculated $\chi$ using the numerical renormalization group method 
with discretization parameter $\Lambda=2$ following 
Krishna-Murthy et al. \cite{Krishnamurthy 80}. At high temperatures, complete diagonalization 
of the logarithmically discretized 5-site model yields essentially exact results (continuous 
lines), as no low energy information is needed in that case. 
The results for fixed hybridization $\Delta=10$ and half bandwidth $D=1$ are shown in 
\mbox{figure \ref{fig:chi_impurity}}. 
At large $T$, the almost free orbital leads to $\chi_{\mathrm{imp}} = \frac{1}{8T}$.
Upon lowering the temperature, we find a characteristic dependence on the value 
of the interaction.
For $U \lesssim 2 \Delta$ we see a crossover to a negative $\chi_{\mathrm{imp}}$, which is due 
to the loss of spectral weight at low frequencies as a consequence of hybridization.
After a characteristic minimum at finite temperature the susceptibility saturates 
at a negative value as $T\to 0$. 
For $U\gtrsim 2 \Delta$, we recover the usual positive $\chi_{\mathrm{imp}}$ which 
is strongly enhanced at large $U/\Delta$ by the Kondo effect defining a new exponentially 
small energy scale $T_K$. 
In spite of these different types of behaviour (depending on the interaction strength), 
the impurity contribution to the density of states at the Fermi energy is negative  
(\ref{reduction}) for \emph{any} $U$. We therefore conclude that only in the 
weakly interacting case $\Delta\rho(\epsilon_F)$ yields the thermodynamic properties 
of the model. In the strong coupling regime, the susceptibility enhancement 
is determined by the many-body resonance in $\rho_d(\omega)$ which -- 
similar to the wide band case -- has to 
be interpreted as a quasiparticle peak with a large effective mass.
The low-temperature behaviour
of the system for strong correlations is governed by many-particle
excitations which are not contained in $\Delta\rho(\omega)$ and,
in fact, completely predominate over the reduction in the single-particle DOS
$\Delta\rho(\epsilon_F)<0$. 

\section{Spectral density and skeleton expansion}
Next, we discuss the spectral density of narrow band Anderson systems.
We calculate the impurity orbital Green's function at zero temperature 
using the Lanczos method as implemented by Krauth \cite{Georges 96}. 
We fix the values of the interaction $U$ and the hybridization $\Delta$ and 
then successively reduce the bandwidth, thus taking the limit $D\to 0$. For each set of parameters 
we calculate the spectral function $\rho_d$, using an Anderson star with $11+1$ sites. For not too 
large $U/\Delta$ the spectral density is found to be 
only weakly dependent on the number of sites. 
%
%
%
%An example of the results
%is given in \mbox{figure \ref{fig:spectrum_widerange}}.
%
%\begin{figure}[t]
%\begin{center}
%\epsfig{file=plot2.eps,width=\linewidth}
%\epsfig{file=plot_rho_N12_U2.0_w1.0e-3_Delta1.0_b1.e-3.eps,width=0.93\linewidth}
%\end{center}
%\caption{\label{fig:spectrum_widerange}Impurity spectral function for an interaction $U=2.0$, hybridization 
%$\Delta=1.0$ and a bandwidth $D=10^{-4}$, evaluated with a broadening parameter $b=1.0\times10^{-3}$.}
%\end{figure}
%

In the limit of small bandwidth at any finite value of the ratio $U/\Delta$, 
we find a three-peak structure 
consisting of the atomic levels at $\omega=\pm U/2$ containing almost the full spectral weight 
and a central quasiparticle peak of width $\sim D$. 
Apart from numerical broadening effects there is no spectral weight between the peaks.
This gives rise to resonances in the imaginary part of the self-energy, which can be seen in the 
following way \cite{Kehrein 98,Zhang 93}:
As a consequence of Dyson's equation, for values of $\omega$ inside the gap 
(where spectral function and hybridization vanish) the self-energy 
$\Sigma(\omega^+) = K(\omega) - i J(\omega)$
is given by  
\beq
\Sigma(\omega^+)= \omega - \Lambda(\omega) - \frac{1}{\Gamma(\omega) - i0^+} \ ,
\eeq
where we have defined 
$\Gamma(\omega) \equiv {\cal P} \int d\epsilon \frac{\rho_d(\epsilon)}{\omega - \epsilon}$. 
The imaginary part of the self-energy therefore has the form
\beq
J(\omega) = \pi \delta\left( \Gamma(\omega) \right) \ .
\eeq  
For the spectral density found here $\Gamma(\omega)$ has zeroes at energies 
$\epsilon \sim \pm\sqrt{D}$ and this leads to $\delta$-functions in
$J(\omega)$ inside the gap as shown above.

As argued previously\cite{Kehrein 98}, these resonances are incompatible with the skeleton expansion, i.e.\ the 
self-consistent perturbation theory in $U$. Within this expansion, the full propagator $G_{dd}(\omega)$ is inserted into 
every diagram contributing to $J(\omega)$. As $G_{dd}(\omega)$  possesses spectral weight only on the small energy scale 
$D \ll \sqrt{D}$, there is no possibility to generate the resonances at $\pm \sqrt{D}$. We therefore conclude 
that in our model already at small but finite bandwidth $D$ the skeleton expansion breaks 
down at energies of the order $\sqrt{D}$ (strictly speaking $\sim\sqrt{\Delta D}$, but note   
that here we have chosen $\Delta=1$ dimensionless).

\begin{figure}[t]
\begin{center}
\epsfig{file=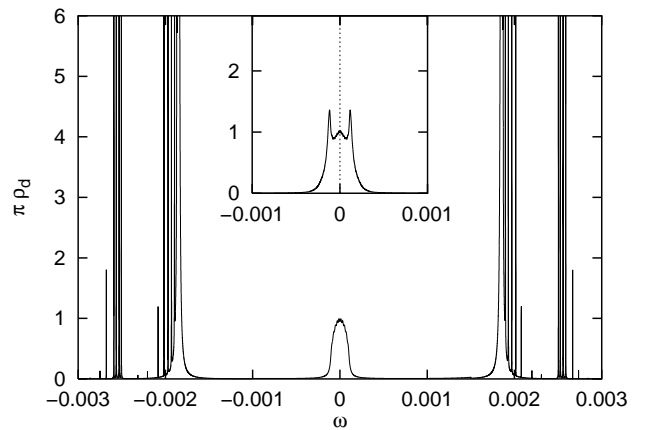,width=0.98\linewidth}
\end{center}
\caption{\label{fig:central_peak}Central peak in the impurity density of states for different 
interactions $U=0.2$ and $4.0$ (inset), hybridization $\Delta=1.0$ and bandwidth $D=10^{-4}$. 
We have used a Lorentzian broadening of $b=1.5\times 10^{-5}$ at the quasiparticle resonance 
and $b=5\times10^{-8}$ for the poles at higher frequencies.
The atomic levels
at $\pm U/2$ carrying most of the spectral weight are not shown. At the Fermi edge, 
$\pi \rho_d(0) = 1.00\pm0.02$ for both values of $U$.}
\end{figure}

As a measure of the convergence of the expansion at lower energies we take the Fermi liquid properties 
of the Anderson model, especially the ``pinning'' of the density of states at its noninteracting 
value (\ref{pinning}). This is equivalent to the vanishing of the imaginary part of the 
self energy at the Fermi level. This property has been proven for a general class of systems 
by Luttinger\cite{Luttinger 61} using the skeleton expansion to all orders. A proof for the (flat band) Anderson 
model within unrenormalized perturbation theory was given by Yamada and Yosida\cite{Yamada 75}.
In order to verify the pinning we focus on the central peak of $\rho_d$  
and compare different, not too strong interactions at constant broadening.
The pinning of $\rho_d(0)$ in the limit $D\to 0$ is evident from figure 
\ref{fig:central_peak}. 
At even larger interactions (not shown here), a narrow Kondo resonance develops inside the band, which 
cannot be resolved well on a small cluster. From the above, though, 
we do not expect deviations from Luttinger's theorem for any interaction.
As can also be seen in figure \ref{fig:central_peak}, the 
``quasiparticle'' resonance in $\rho_d(\omega)$
has an internal structure itself (including bound states outside the band for small $U/\Delta$).   
In the limit $D\to 0$ this can be described  by a scaling function 
\beq \label{scaling}
\rho_d(\omega) = f(\omega/D) \ ,
\eeq
where $f$ is independent of $D$. The $\delta$-peaks in figure \ref{fig:central_peak}, dominant at 
small $U$, can be understood by approximating the Hamiltonian as a ``zero bandwidth'' Anderson 
model \cite{Hewson 93} where the impurity couples to one single orbital and the hybridization is given by 
a $\delta$-function carrying the integrated weight $\int d\epsilon \Delta(\epsilon)$.
The effect of the continuous hybridization band is to generate spectral density close to $\omega=0$ 
and to create sidebands also visible in figure \ref{fig:central_peak}.
For larger values of $U$ the weight of the bound states decreases (from the zero bandwidth 
model we expect a decrease  $\sim 1/U^2$) and they numerically merge with the continuum at small $\omega$.

For other hybridization functions in the narrow band limit
(\ref{def_narrowband}) the main conclusions of the above analysis
remain unaffected. A detailed numerical study is however difficult
due to the limited number of orbitals that can be taken into account
using exact diagonalization. For example in a ``narrow'' band with 
band tails extending to $\pm\infty$ but 
$\left.\frac{\partial\Lambda}{\partial\omega}\right|_{\omega=0}\gg 1$,
the bound states for $U=0$ become sharp resonances. Likewise the
imaginary part of the self-energy then contains sharp resonances 
at $\pm O(\sqrt{D})$ instead of $\delta$-functions. Still the
resonances contain the same spectral weight as these $\delta$-functions
and the breakdown of the skeleton expansion occurs in the same manner
\cite{Kehrein 98}.

\section{Conclusion}
In this paper we have studied the Anderson impurity model in the narrow band 
limit (\ref{def_narrowband}) using numerical methods. We have found two new
features as compared to the usual wide band limit. i) One observes a crossover
in the impurity contribution to the susceptibility: For small interactions the
impurity reduces the total susceptibility, for large interactions the impurity increases
it. In fact, the same behaviour is also found for the specific heat, though
this has not been discussed explicitly here. This crossover is in contrast 
to the observation
that the impurity contribution to the total density of states at the
Fermi level is \emph{always} negative. 
This quantity does therefore not determine the 
thermodynamic properties of the system at large interactions.
ii) Holding $U$ and $\Delta$ fixed, the skeleton expansion breaks down for 
sufficiently small (but still finite) bandwidth~$D$. The breakdown occurs at
energies of order $\sqrt{D}$ and larger, while for smaller energies no problems
can be found. This shows that this specific breakdown of the skeleton expansion
is a generic feature of narrow band Anderson impurity systems \cite{Lange 98}. 
On the other hand, the skeleton
expansion is an essential tool for deriving the locality of the self-energy
in the dynamical mean field theory in the Fermi liquid phase \cite{Metzner 89,Mueller-Hartmann 89}. 
Its convergence also provides a sufficient condition for the analytic
continuation to the noninteracting Hubbard model. 
At present, the question of the correct description of the Mott-Hubbard
transition in large dimensions is under debate \cite{Kehrein 98,Nozieres 98,Gebhard 98}:
While NRG simulations \cite{Bulla 98a,Bulla 98b} at $T=0$ seem to 
indicate a ``preformed gap'', no coexistence between metallic and insulating solutions 
is found in finite temperature Quantum Monte Carlo calculations \cite{Schlipf 98}.
In any case, the analysis presented above shows that for the preformed gap scenario  
(which naturally leads to an effective action governed by a narrow band system
in the sense of (\ref{def_narrowband})) one has to address the question whether such systems
can be related to the original Hubbard model in large dimensions.

The authors would like to thank D.\ Vollhardt, F.\ Gebhard, P.G.J.\ van Dongen, 
V.\ Zlati\'{c}, T.A.\ Costi and R.\ Bulla for valuable discussions.

\end{document}